\documentclass[amsmath,amssymb,aps,prl,twocolumn,floatfix,superscriptaddress]{revtex4-1}

\usepackage{graphicx,dcolumn,bm,color,multirow,float}

\begin{document}

\title {Magnetic Ordering, Anomalous Lifshitz Transition and Topological Grain Boundaries in Two-Dimensional Biphenylene Network}

\author{Young-Woo Son}
\email{hand@kias.re.kr}
\affiliation{
Korea Institute for Advanced Study, 
Seoul 02455, Korea
}
\author{Hosub Jin}
\email{hsjin@unist.ac.kr}
\affiliation{
Department of Physics, Ulsan National Institute of Science and Technology,
Ulsan 44919, Korea
}
\affiliation{
Korea Institute for Advanced Study, 
Seoul 02455, Korea
}

\author{Sejoong Kim}
\affiliation{University of Science and Technology (UST), Gajeong-ro 217, Daejeon 34113, Korea}
\date{\today}

\begin{abstract}
We study electronic properties of a new planar carbon crystal formed through networking biphenylene molecules. Novel electronic features among carbon materials such as zone-center saddle point and peculiar type-II Dirac fermionic states are shown to exist in the low energy electronic spectrum. The type-II state here has a nearly flat branch and is close to a transition to type-I. Possible magnetic instabilities related with low energy bands are discussed. Furthermore, with a moderate uniaxial strain, a pair of Dirac points merge with the zone center saddle point, realizing concurrent Lifshitz transitions of van Hove singularity as well as pair annihilation of the Dirac fermions. A new effective Hamiltonian encompassing all distinctive low energy states is constructed, revealing a finite winding number of the pseudo-spin texture around the Dirac point, quantized Zak phases, and topological grain boundary states. 
\end{abstract}

\maketitle

Graphitic materials have several notable electronic structures such as the Dirac fermionic states in a wide range of energy-momentum space~\cite{Wallace1947PR,Charlier2007RMP,CastroNeto2009RMP}.
These properties are realizable partly owing to the large hopping energy,  hexagonal crystal symmetry and negligible spin-orbit interaction (SOI)~\cite{Charlier2007RMP, CastroNeto2009RMP}.
Recently, a new network of carbon atoms with nonhexagonal rings has been synthesized~\cite{Fan2021Science}. The new planar 
carbon allotrope or biphenylene network (BPN) was grown into the ribbon shape on the Au(111) surface of which electronic properties are dependent of its width~\cite{Fan2021Science}. 
Although they are still one-dimensional, its extension to two-dimensional (2D) form like graphene is surely expected. 
Considering that carbon materials have been fruitful playgrounds for many fundamental physical concepts~\cite{Charlier2007RMP,CastroNeto2009RMP}, we anticipate that BPN could also have potentials to 
demonstrate new physical phenomena.

In this work, we show that, unlike any other carbon material, a pristine 2D BPN has a saddle shaped energy band at the Brillouin zone (BZ) center and peculiar type-II Dirac fermion. The type-II state here has a nearly flat branch and is shown to support topologically protected grain boundary state.
With moderate uniaxial strains, two Dirac cones are driven to merge with the saddle point, thus invoking simultaneous Lifshitz transitions for the van Hove singularity (vHS) and type-II Dirac fermions.
Moreover, including effects of Coulomb interactions, the ground state turns out to be ferrimagnetic.
Considering control techniques developed for 2D materials~\cite{Novoselov2016Science}, we expect the new carbon allotrope~\cite{Fan2021Science} could be one of quintessential materials to study interplays between topological and correlated properties.

We studied electronic properties of BPN based on first-principles methods~\cite{QE2009JPCM} and effective models. 
The generalized gradient approximation (GGA)~\cite{GGA1996PRL} was used for the exchange-correlation functional.
A plane-wave basis set with a cutoff energy of 100 Ry is adopted and the ultrasoft pseudopotential~\cite{GVRB2014CM} is used. 
To consider Coulomb interaction beyond the GGA, we used a newly developed DFT+$U$+$V$ method exploiting the self-consistent evaluation~\cite{ACBN2015PRX,Lee2020PRR,Rubio2020PRB} of the on- and inter-site Hubbard interactions ($U$ and $V$)~\cite{Cococcioni2010JPCM}. 
We consider $V$ between the nearest neighbors (n.n.), that was confirmed to achieve desired convergence~\cite{Lee2020PRR}.
The method enable us 
to include Coulomb interaction in a similar level of $GW$ approximations with a significantly reduced computational cost~\cite{Lee2020PRR,Rubio2020PRB,Cococcioni2010JPCM,Timrov2021PRB}. We note that this method captures the effects of Coulomb interaction in 2D crystals such as black phosphorous~\cite{Lee2020PRR} and graphene~\cite{Rubio2020PRB} very well.

\begin{figure}[t]
	\begin{center}
		\includegraphics[width=1\columnwidth]{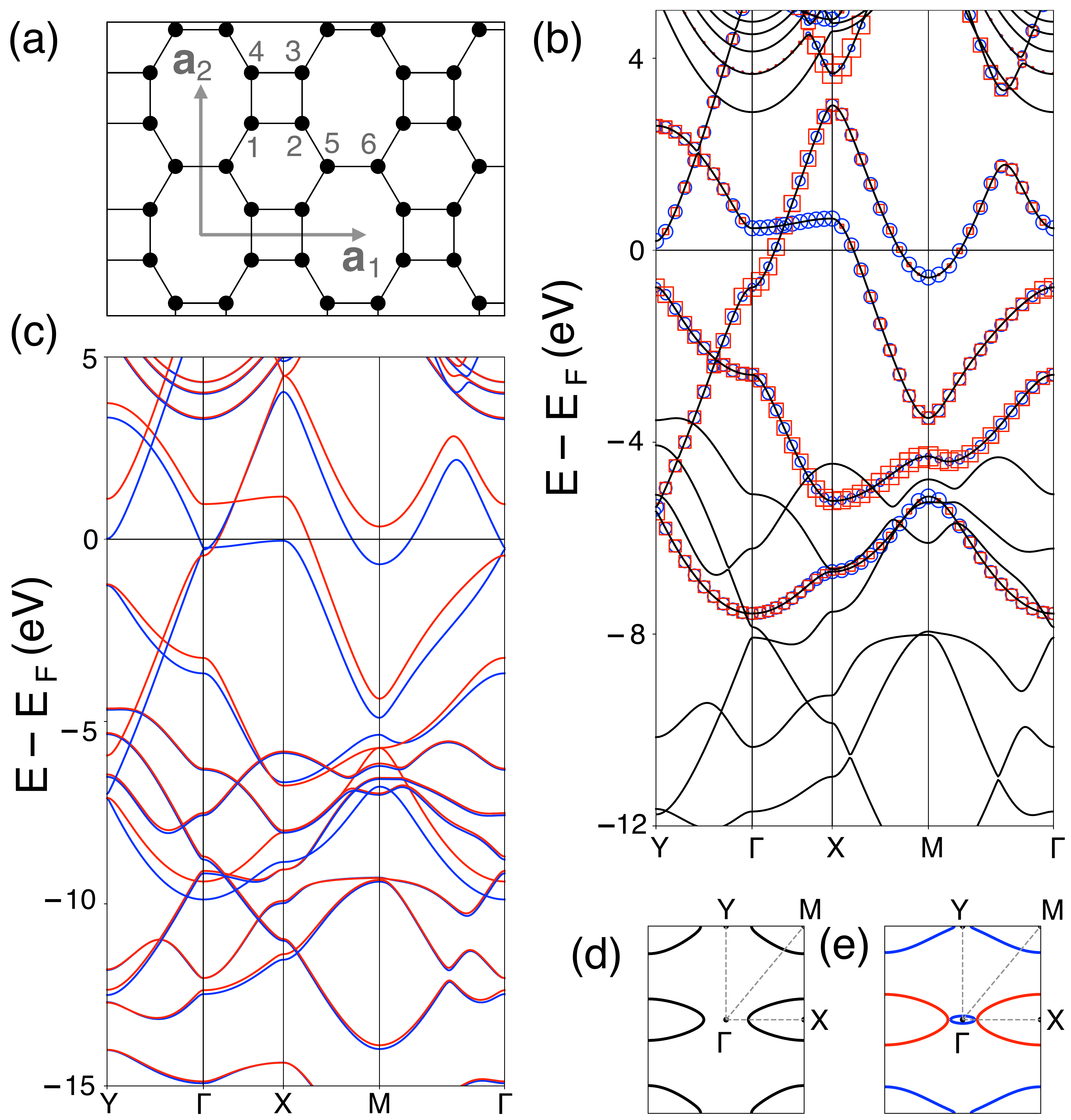}
	\end{center}
	\caption{(a) A single layer BPN lattice. Black dots denote carbon atoms. 
	For vectors (${\bf d}_{ij}$) connecting the nearest neighbors $i$ and $j$ (=$1,\cdots,6$), $|{\bf d}_{12}|=|{\bf d}_{23}|=1.46$, $|{\bf d}_{56}|=1.45$ and $|{\bf d}_{25}|=1.41$~\AA, respectively. Angles of $\theta_{ijk}$ between ${\bf d}_{ji}$ and ${\bf d}_{jk}$ are $\theta_{123}=\theta_{234}=90^\circ$ and $\theta_{256}=124.9^\circ$.
	${\bf a}_1=a_1\hat{x}$ and ${\bf a}_2=a_2\hat{y}$ are unitcell vectors where $a_{1(2)}=4.51 (3.76)$~\AA. 
	(b) DFT-GGA band structures. Sizes of blue circles and red rectangles are proportional to the energy-momentum resolved density of states per atom projected on the $p_z$ orbitals belonging to square and dimer units in (a), respectively. Group velocities for linear bands around the Dirac points are 9.6$\times 10^3$ and 6.1$\times10^2$ m/s, respectively
	(c) Band structures obtained by DFT+$U$+$V$ method. The red and blue lines denote the bands with opposite spin orientations.
(d) and (e) Fermi surfaces of energy bands in (b) and (c), respectively. 
	}
	\label{fig:basic}
\end{figure}

The 2D BPN lattice in Fig.~\ref{fig:basic}(a) has the point group $D_{2h}$ with 4-, 6- and 8-membered rings. It can be regarded as an effective lattice where a square shaped cluster consisting of atoms indexed by $1,\cdots,4$ in Fig.~\ref{fig:basic} (a) forms a rectangular lattice and a dimer of atoms indexed by 5 and 6 is its basis. By assigning an effective molecular orbital with specific angular momentum for each unit as a basis set, we can describe the low energy electronic states of the BPN lattice. This may offer a unique chance to support interesting
correlated  states~\cite{Guinea1997PRL, Furukawa1998PRL, Honerkamp2001PRL, Yu2001PRB} as well as symmetry protected energy bands~\cite{Young2015PRL, Kim2015PRL, Bernevig2015Nature, Shuyun2017NatComm, Ashvin2018RMP}.

All structures are fully relaxed 
and dynamical stability of free standing BPN is confirmed by checking its stable phonon dispersion using density functional perturbation theory~\cite{DFPT2001RMP,QE2009JPCM}, agreeing with a recent study~\cite{Jain2021Carbon}.
Detailed information on crystal structure is in the caption of Fig.~\ref{fig:basic}.
Although not all the local bonds are ideal hexagonal $sp^2$ bonds, the states are completely decomposed into the $\pi$ and $\sigma$ bonds, as shown by projected local density of states (DOS) in Fig.~\ref{fig:basic} (b). 
The non-ideal local $\sigma$ bonds manifest themselves in different charge states depending on the atomic positions. 
Our calculation shows that the L\"owdin charge of the square unit in Fig.~\ref{fig:basic} (a) is smaller by 0.13$|e|$ ($e$ is an electron charge) per atom than one in the dimer unit, implying the higher site energy of the former than the latter.

We first highlight two novel features in energy bands of a single layer BPN. 
The saddle shaped band intersects with the almost flat band near $\frac{2}{5}\Gamma\text{X}$, forming a type-II Dirac state~\cite{Bernevig2015Nature, Shuyun2017NatComm, Ashvin2018RMP}.
 As shown in Fig.~\ref{fig:basic} (b), the nearly flat band is completely projected on the $p_z$-orbitals localized within the square unit.
The almost flat dispersion indicates that the state is close to a transition between the  type-I and II Dirac states~\cite{Volovik2017JLTP,McCormick2017PRB}.
The vHS at the BZ center is also uncommon since it is usually at the zone boundary~\cite{Guinea1997PRL, Furukawa1998PRL, Honerkamp2001PRL, Yu2001PRB,Kim2021PRB}.
These topological bands should be maintained under strong Coulomb interactions unless ground state symmetries are broken. Like other graphitic materials~\cite{Charlier2007RMP, CastroNeto2009RMP}, however, the Coulomb interactions will alter energies and slopes of DFT-GGA bands.

\begin{figure}[!t]
	\begin{center}
		\includegraphics[width=1.0\columnwidth]{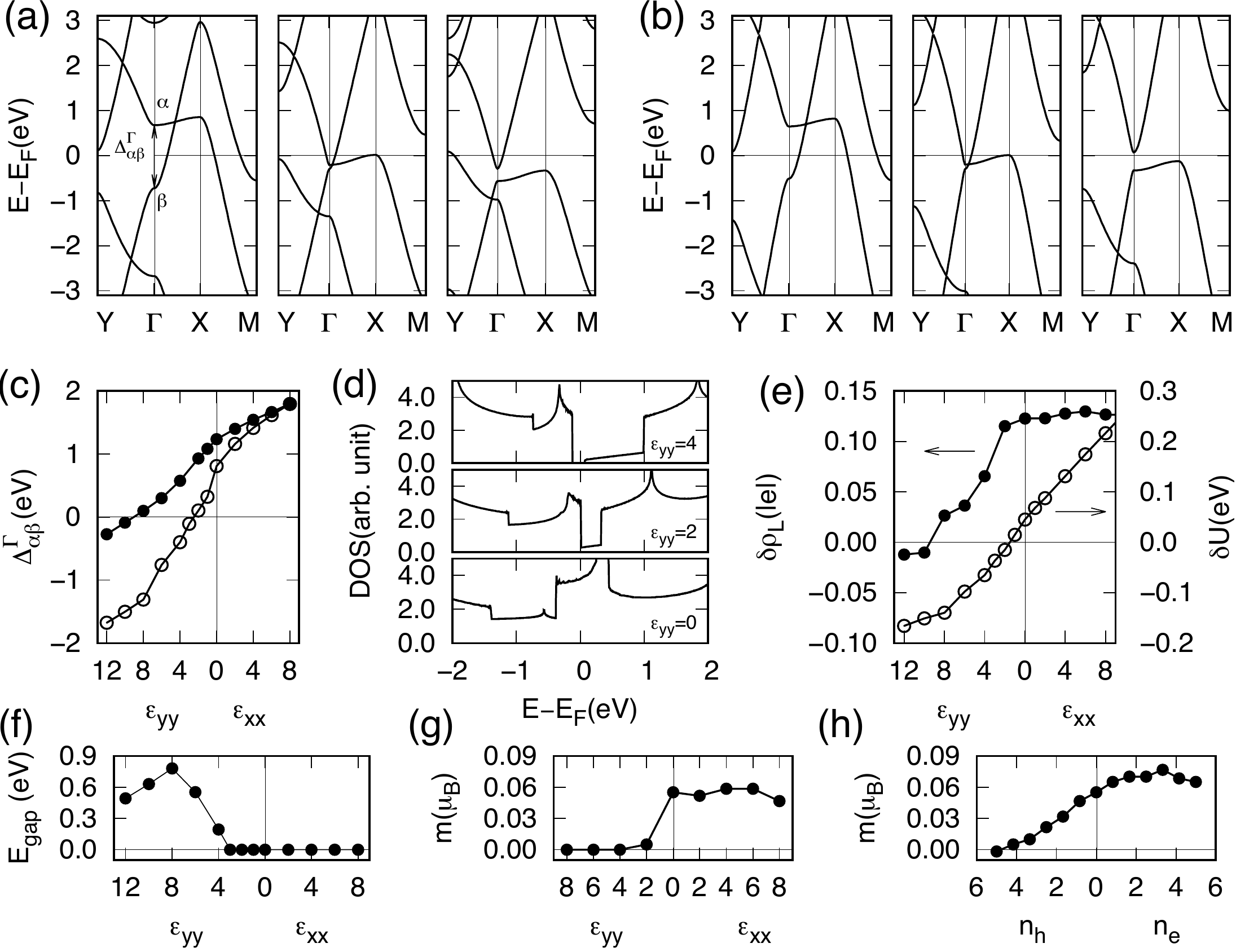}
	\end{center}
	\caption{Energy bands under strains obtained with (a) DFT-GGA and (b) DFT+$U$+$V$, respectively. From the left to right panels of (a), the bands with $\varepsilon_{xx}$ of 2 (in unit of \%) and $\varepsilon_{yy}=$ 8 and 12, respectively. Two bands near $E_F$ are labeled by $\alpha$ and  $\beta$. 
	From the left to right panels of (b), $\varepsilon_{xx}=2$, $\varepsilon_{yy}=2$ and 4, respectively. 
	(c) $\Delta^\Gamma_{\alpha\beta}$ as a function of strains using DFT-GGA  (filled circles)
	and DFT+$U$+$V$ (empty circles), respectively.
	(d) Strain-induced variations of density of states (DOS) using DFT+$U$+$V$.
	(e) L\"owdin charge ($\delta\rho_L$) as well as onsite energy ($\delta U$) difference  as a function of strains. 
	Here, $\delta\rho_L\equiv \rho_\mathcal{S}-\rho_\mathcal{D}$ where $\rho_\mathcal{S(D)}$ is the L\"owdin charge per atom in the square (dimer) unit and 
	$\delta U\equiv U_\mathcal{S}-U_\mathcal{D}$ where $U_\mathcal{S(D)}$ is the self-consistently evaluated on-site Hubbard repulsion for $p_z$-orbital belonging to square (dimer) unit. 
	(f) Indirect band gaps with DFT+$U$+$V$ as a function of applied strains. 
	 Variation of magnetic moment per atom ($m(\mu_B)$) as a function of strain (g) and doping (h).
	 $n_{e(h)}$ denotes electron (hole) doping per atom in a unit of $\mp 10^{-2} |e|$ where $e$ is an electron charge.
	}
	\label{fig:strain}
\end{figure}

Including Coulomb interactions beyond the GGA, we obtain a ferrimagnetic ground state. Our self-consistent evaluation of Hubbard interactions are $U=5.93~(5.98)$ eV and $V=3.05~(3.09)$ eV for $p_z$-orbitals in the square (dimer) unit. These values are comparable to $U$ and $V$ of graphene~\cite{Wehling2011PRL, Rubio2020PRB}. Unlike graphene where all atoms have the same Hubbard interactions, atoms belonging to different units of BPN lattice have different values.
Moreover, diverging DOS near $E_F$ can invoke large exchange splitting. Our {\it ab initio} calculation indeed obtains a net magnetic moment of $0.33~  \mu_B$ per unit cell where a moment per atom belonging to square (dimer) unit is $0.12~(-0.07)~ \mu_B$.
The total energy for the magnetic state is lower than nonmangetic one by 14.2 meV per atom. 
As shown in Fig.~\ref{fig:basic} (c), spin resolved band structures have four crossing points where two of them are quite close to the zone center and $E_F$. 
The energy level of vHS is also closer to the $E_F$ than DFT-GGA, 
as also can be seen in 2D Fermi surfaces for each case shown in Figs.~\ref{fig:basic} (d) and (e), respectively.
We note that our result is consistent with several theoretical studies~\cite{Guinea1997PRL, Furukawa1998PRL, Honerkamp2001PRL, Yu2001PRB, Lin2015PRB} on effects of interaction near vHS. While the strong repulsion within a single vHS alone may not be so effective in realizing spin ordering~\cite{Guinea1997PRL}, 
the half-filled vHS could realize a magnetic instability within the Stoner limit~\cite{Lin2015PRB}.
We also note that the computed work function of 4.24 eV with extended Hubbard interactions slightly differ from 4.32 eV with DFT-GGA.

As demonstrated in the other 2D single layer magnets~\cite{Gong2017Nature, Kim2019NatComm},
our magnetic state may not be so stable in the limit of single layer owing to intrinsic fluctuations with negligible SOI here. The magnetic state may stabilize by an enhanced anisotropy from external fields or substrate. So, hereafter, we include the $U$ and $V$ without magnetism to take into account the effects of Coulomb interactions on band energies and group velocities like previous calculations for nonmagnetic materials~\cite{Lee2020PRR, Rubio2020PRB}.

Having discussed unique low energy bands, now we turn to their control. 
Like graphene~\cite{Pereira2009PRL, Choi2010PRB} 
or black phosphorous~\cite{baik2015NL,Doh2017Mat},
strain can alter electronic and topological properties of a single layer BPN.
We consider uniaxial strains of $\varepsilon_{xx}~(\varepsilon_{yy})$ along $x(y)$-direction. With increasing $\varepsilon_{xx}(\varepsilon_{yy})$, Dirac points moves to the BZ boundary (center) [Figs.~\ref{fig:strain} (a)].  The energy difference ($\Delta^\Gamma_{\alpha\beta}$ in Fig.~\ref{fig:strain} (a)) between the nearly flat and saddle shaped bands at the $\Gamma$ point also increases (decreases) as $\varepsilon_{xx}~(\varepsilon_{yy})$ increases as shown in Fig.~\ref{fig:strain} (c).
With increasing $\varepsilon_{yy}$, the Dirac point located on $\Gamma$X and its time reversal partner approach each other and eventually merge together at the $\Gamma$-point.
Unlike typical merging~\cite{Murakami2007PRB,Son2011PRB} 
or generation~\cite{baik2015NL,Doh2017Mat} of a pair of Dirac cones, 
they annihilate each other at the saddle point as shown in Figs.~\ref{fig:strain} (a) and (c).
Thus, its DOS at the transition diverges unlike the usual semi-Dirac cone at the critical point~\cite{Dietl2008PRL, Son2011PRB}
as shown in Fig.~\ref{fig:strain} (d).
The sign change of $\Delta^\Gamma_{\alpha\beta}$ near $\varepsilon_{yy}\simeq 9\%$ in Fig.~\ref{fig:strain} (c) signals gap opening at the zone center, occurring the Lifshitz transition.
This transition is also accompanied with charge transfer between the square and dimer units. As shown in Fig.~\ref{fig:strain}(e), the charge difference between two units ($\delta\rho_L$) changes its sign at the $\varepsilon_{yy}$ for gap opening. Thus, phenomenologically, we correlate the site energy or local charge reversal between two units with the topological transition. 

\begin{figure}[t]
	\begin{center}
		\includegraphics[width=0.8\columnwidth]{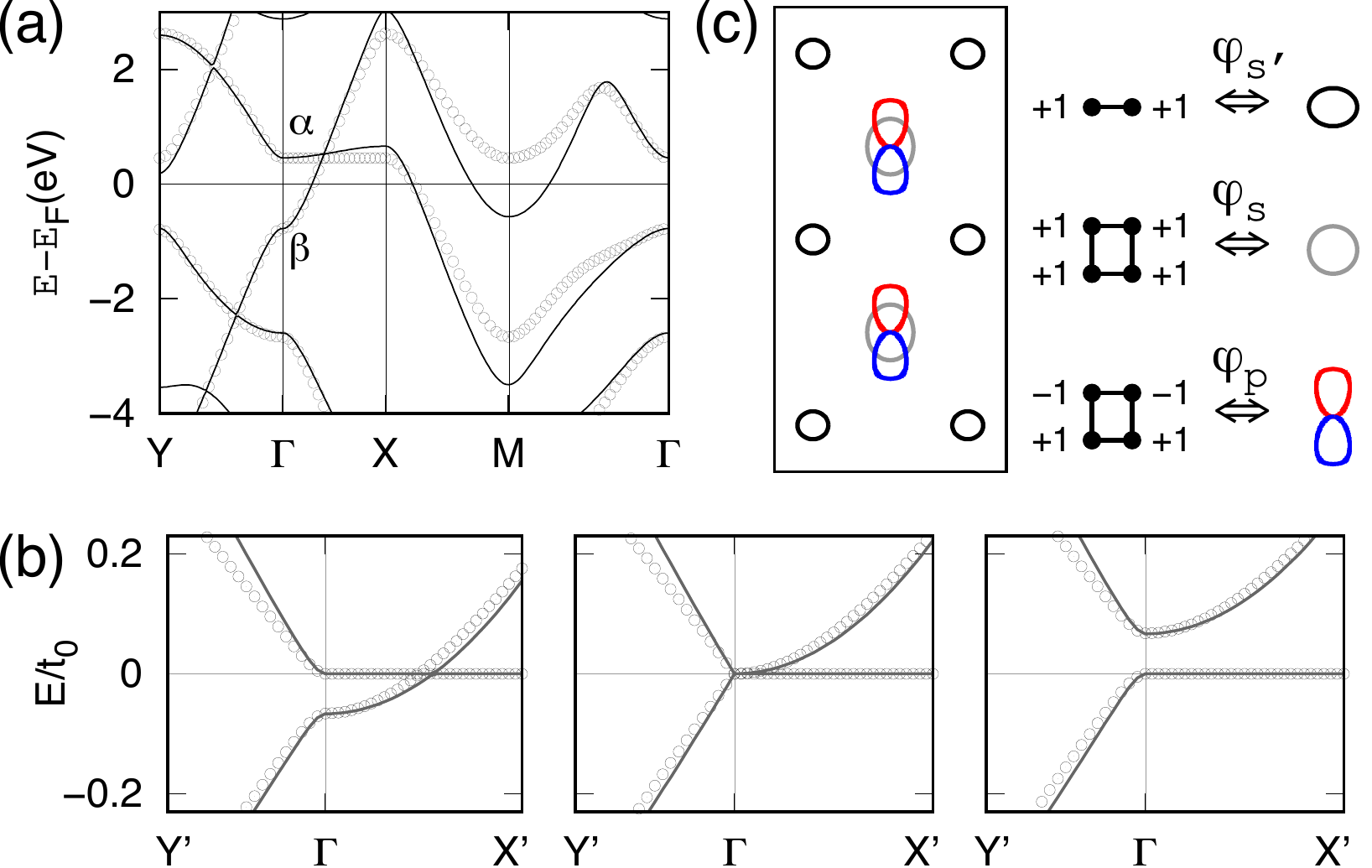}
	\end{center}
	\caption{(a) Circles denote bands from TB Hamiltonian Eq.~\ref{eq:tbH} 
	and lines from DFT-GGA. 
	(b) Circles represent TB bands from Eq.~\ref{eq:tbH} 
	with $t_d=0$ and $\text{Y}'=0.2\text{Y}$, 
	$\text{X}'=0.3\text{X}$. From the left to right panels, $\epsilon_D/t_0 =-0.1$, $0$ and $+0.1$, respectively. Gray lines are corresponding effective bands from Eq.~\ref{eq:effH} based on the effective model described in (c).
	(c) An effective orbital model on the BPN lattice. The dimer and square units with same phase are regarded as effective $s$-orbitals denoted by $\varphi_{s'}$ and $\varphi_{s}$, respectively while the square unit with $\pi$-phase change along $y$-direction as an effective $p_y$-orbital denoted by $\varphi_p$ where blue and red lines denote $\pm 1$ phase, respectively.
	}
	\label{fig:tb_band}
\end{figure}

With $U$ and $V$, all the variations by strain discussed above still hold. The band width increases by including inter-site Hubbard interaction of $V$~\cite{Lee2020PRR, Rubio2020PRB, Cococcioni2010JPCM} so that the Dirac points are closer to $E_F$ and $\Gamma$-point than ones from DFT-GGA, resulting in a smaller critical strain for the topological transition. So, as shown in Figs.~\ref{fig:strain} (b) and (c), the Lifshitz transition achieves with $\varepsilon_{yy}\simeq 2.5\%$ for DFT+$U$+$V$ bands while $\varepsilon_{yy}\simeq 9\% $ for DFT-GGA bands as also shown in sign change of $\Delta^\Gamma_{\alpha\beta}$ in Fig.~\ref{fig:strain} (c). Like $\delta\rho_L$ for DFT-GGA bands, the difference of on-site Hubbard repulsion between two units ($\delta U$) also changes its sign near the transition as shown in Fig.~\ref{fig:strain} (e). 
Unlike semimetallic nature of strained BPN within DFT-GGA calculations, we obtain indirect band gaps with the Hubbard interactions when $\varepsilon_{yy}> 3\%$ [Fig.~\ref{fig:strain} (f)] owing to enhanced band repulsion by $U$. So, the annihilation of Dirac cones by strain induces metal-insulator transition with the maximum gap of 0.78 eV at $\varepsilon_{yy}=8\%$.

Having described variations in electronic energy bands, we briefly discuss control of magnetic moments under external perturbations. 
As shown in Fig.~\ref{fig:strain} (g), the magnetic moment quickly disappears as the energy gap develops with $\varepsilon_{yy}$ while it maintains the value with $\varepsilon_{xx}$. The variation of magnetic moments also can be achieved with external dopings. 
In Fig.~\ref{fig:strain} (h), it is shown that the moment decreases gradually with increasing hole doping while it remains with electronc doping since the large difference between hole and electron side density of states shown in Fig.~\ref{fig:strain} (d).

Hereafter, We discuss origins for the novel bands and their topological consequences with simpler models. All features found in our {\it ab initio} bands can be retrieved by tight-binding (TB) approximations with the $p_z$-orbitals. 
Since the fine-tuned hopping energy depending on n.n. distances only changes the minute details, we set them to be a same value of $-t_0$. 
A hopping energy of $-t_d$ along diagonal directions in the square unit is included to obtain realistic hopping energies.
To account for $\delta\rho_L$ or $\delta U$ in Fig.~\ref{fig:strain} (e), 
the site energy of $\epsilon_D$ for the atoms in the dimer unit is indispensable. 
With these parameters, a TB Hamiltonian can be written as
\begin{equation}
    {\mathcal H}_\text{TB}=
    -t_0 \sum_{\langle i,j\rangle}c^\dagger_i c_j
    -t_d \sum_
    {\langle\langle i,j\rangle\rangle \in {\mathcal S}}c^\dagger_i c_j
   + \epsilon_D \sum_{i\in {\mathcal D}}c^\dagger_i c_i +(\text{c.c.}),
    \label{eq:tbH}
\end{equation}
where $\mathcal{S(D)}$ denotes a set of atoms belonging to the square (dimer) units and $\langle\cdots\rangle$ and $\langle\langle\cdots\rangle\rangle$ indicate the n.n. and next n.n. pairs, respectively.
 Comparing TB bands with DFT-GGA ones, 
 we obtained $t_0=2.59$ eV, $t_d=0.45$ eV and $\epsilon_D=-0.95$ eV, respectively, and the TB bands agree with our DFT-GGA bands [Fig.~\ref{fig:tb_band} (a)]. Even without $t_d$, Eq.~\ref{eq:tbH} can reproduce almost same bands and topological characteristics. For such a case, however, the fitted values are unrealistic.   
We also note that a long-range hopping between rectangular units along ${\bf a}_1$ direction makes the flat $\alpha$-band in Fig.~\ref{fig:tb_band} (a) be dispersive to match our DFT-GGA bands better.

The topological transitions also can be described by Eq.~\ref{eq:tbH}. Not by varying hoppings according to strain in detail, the transition can be simply controlled by sign of $\epsilon_D$. As shown in Fig.~\ref{fig:tb_band} (b), the $\epsilon_D$ following the sign of $\delta\rho_L$ or $\delta U$ in Fig.~\ref{fig:strain} (e) realises the transition: with $\epsilon_D <0$, the type-II Dirac point is on the flat band and with $\epsilon_D > 0$, gap opens. When $\epsilon_D=0$ implying the same site energy for two sublattice units, the two tilted Dirac cones and saddle point merge at the zone center resulting in coexistence of critical states and vHS.

We find that wave functions (WFs) at the $\Gamma$-point for the vHS and type-II Dirac states are linear combinations of characteristic states in the square ($\varphi_s$ and $\varphi_p$) and dimer units ($\varphi_{s'}$). The states are expressed with a linear combination of $p_z$ orbitals of atom $i$ ($\phi_i$) where $i=1,\cdots,6$ in Fig.~\ref{fig:basic}(a) such that
$\varphi_{s(p)} =\frac{1}{2}(\phi_1 +\phi_2 \pm\phi_3 \pm\phi_4)$ and $\varphi_{s'}=\frac{1}{\sqrt{2}}(\phi_5+\phi_6)$. 
These WFs for the units can be mapped onto effective orbitals as described in Fig.~\ref{fig:tb_band} (c) 
such that $\varphi_{s(s')}$ is regarded as an effective $s$-orbital and $\varphi_p$ as $p_y$-orbital, respectively.
Then we immediately notice that the hopping channel between neighboring $\varphi_p$ orbitals through $\varphi_{s'}$ along $x$ direction is forbidden, realizing the flat band along $k_x$ direction and localization of its WFs within the square unit as shown in the {\it ab initio} bands [Fig.~\ref{fig:basic}(b)].
We also can explain the zone-center vHS by considering right combination of hopping between $\varphi_p$ and $\varphi_{s'}$ along $y$-direction
and one between $\varphi_s$ and $\varphi_{s'}$ along $x$-direction because their hopping energies should have opposite signs to each other.

The qualitative discussions above can be explicitly expressed by constructing effective models. 
For a basis set of $\Psi=(\varphi_p, \varphi_s,\varphi_{s'})$ using the orbitals in Fig.~\ref{fig:tb_band} (c), 
a Hamiltonian 
for three effective orbitals can be written as
\begin{equation}
\mathcal{H}_\text{orb}({\bf k})=
    \begin{pmatrix}
    \varepsilon_p & 0 & -i f({\bf{k}})  \\
    0 &\varepsilon_s & -g({\bf{k}})\\
     i f({\bf{k}}) & -g({\bf{k}}) & \varepsilon_{s'}
    \end{pmatrix},
    \label{eq:orbH}
\end{equation}
where $\hbar=1$, ${\bf k}=(k_x, k_y)$, $f({\bf{k}})\equiv 4t'\cos\frac{k_x a_1}{2}\sin \frac{k_y a_2}{2}$, 
$g({\bf{k}})\equiv 4t\cos\frac{k_x a_1}{2}\cos\frac{k_y a_2}{2}$, and $\varepsilon_l$ $(l=s,s',p)$ is a site energy for $\varphi_l$. $t(t')$ is a hopping energy between $\varphi_s$ and $\varphi_{s'} (\varphi_p)$.
Without loss of generality, we can set $\varepsilon_p =0$, $t=t'$ and  $\varepsilon_D\equiv\varepsilon_{s'}-4t=\varepsilon_s-4t$, respectively. Then, it can be further reduced to a simpler Hamiltonian by projecting Eq.~\ref{eq:orbH} onto a basis of $\Psi_\text{eff}=(\varphi_p,\tilde{\varphi_s})$ where $\tilde{\varphi_s}=\frac{1}{\sqrt{2}}(\varphi_s+\varphi_{s'})$. 
The resulting ${\mathcal H}_\text{eff}$ near $\Gamma$-point is
\begin{equation}
\mathcal{H}_\text{eff}({\bf k})\simeq
    \begin{pmatrix}
    -\frac{\delta_y}{2m_y}k_y^2 & iv_y k_y\\
    -iv_y k_y & \varepsilon_D+\frac{k_x^2}{2m_x}+\frac{k_y^2}{2m_y}
    \end{pmatrix},
    \label{eq:effH}
\end{equation}
where $\delta_y=\frac{2t}{\varepsilon_D+8t}$, $v_y = \sqrt{2}t a_2$, 
$m_x^{-1}=2t a_1^2$ and $m_y^{-1}=2t a_2^2$.
By matching $\varepsilon_D=-\Delta^\Gamma_{\alpha\beta}\simeq\frac{2}{3}\epsilon_D$, 
and $t=t_0$, the energy bands from Eq.~\ref{eq:effH} reproduce 
the topological transitions for vHS and Dirac states obtained from Eq.~\ref{eq:tbH} 
very well as shown in Fig.~\ref{fig:tb_band}(b).
The critical type-II Dirac Hamiltonian 
can also be obtained by projecting Eq.~\ref{eq:orbH} to $\Psi_\text{eff}$ around the Dirac point of ${\bf k}_D=(\pm k_D,0)$ such as 
\begin{equation}
\mathcal{H}_\text{crit}(q_x\pm k_D, q_y)\simeq
    \pm v_x q_x \sigma_0 \mp v_x q_x\sigma_z-v_y q_y\sigma_y,
    \label{eq:typeII}
\end{equation}
where $k_D =\frac{2}{a_1}\arccos(1+\frac{\varepsilon_D}{4t})$, 
$v_x=\frac{a_1}{2}[-\varepsilon_D(\varepsilon_D+8t)]^{\frac{1}{2}}$, $\sigma_0$ is an identity and $\sigma_{z(y)}$ Pauli matrix.
Eq.~\ref{eq:typeII} indeed satisfies the condition for the transition between types-I and II Dirac states~\cite{Volovik2017JLTP, McCormick2017PRB} 
and ensures the pseudospin structures for nontrivial topological
properties~\cite{Bernevig2015Nature, Shuyun2017NatComm}.

\begin{figure}[t]
	\begin{center}
		\includegraphics[width=0.8\columnwidth]{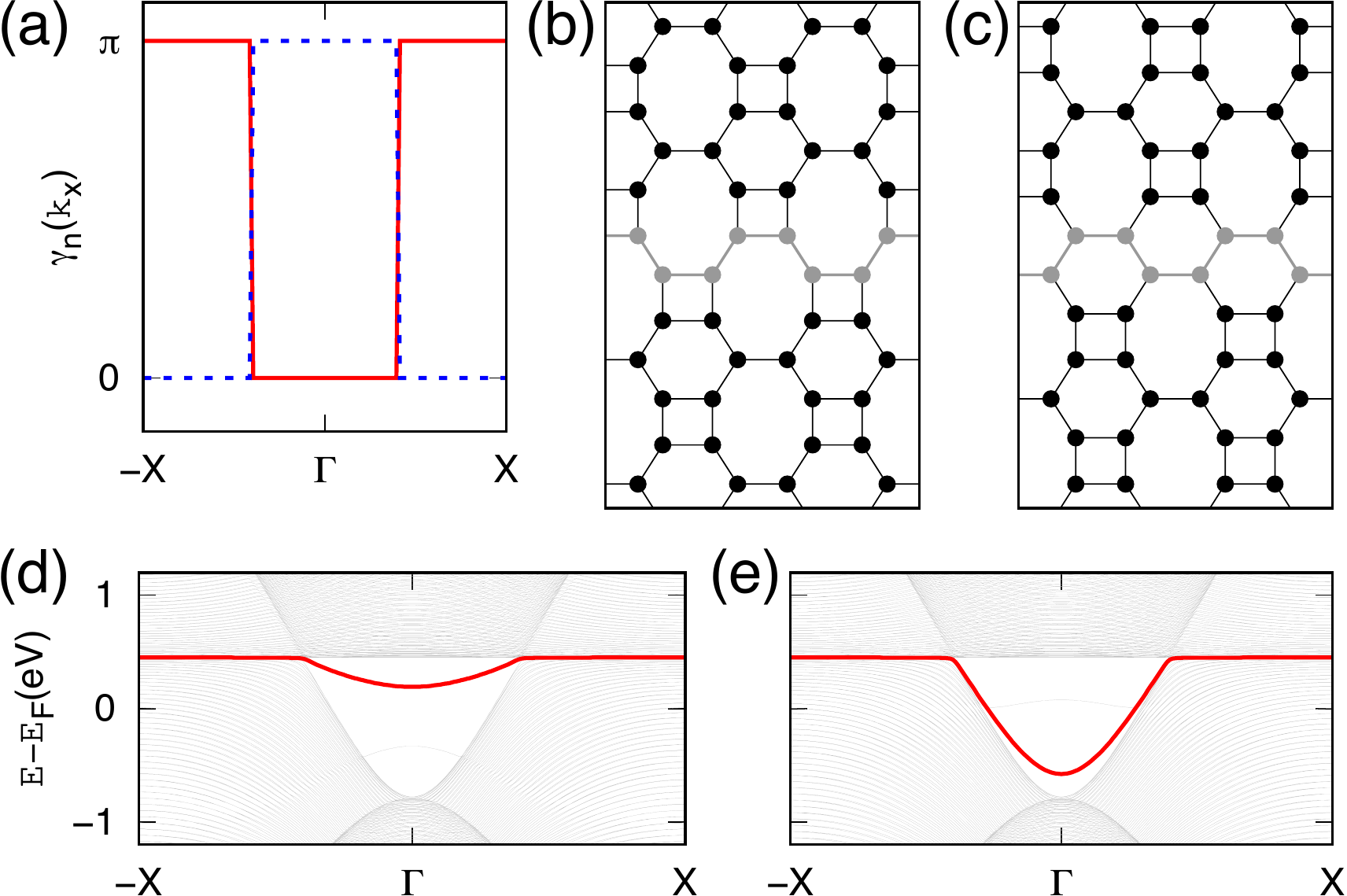}
	\end{center}
	\caption{(a) Calculated Zak phases ($\gamma_n (k_x)$) of $n=\alpha$ (red) and $\beta$ (blue) bands along $k_y$ direction as a function of $k_x$, respectively. (b) and (c) are two different domain boundaries between half-translated unitcells along $x$ direction. Gray color denotes atoms on the boundaries.
	(d) and (e) are projected bands along $k_x$ of the structures in (b) and (c), respectively. The thick red lines denote topologically protected bands.
	}
	\label{fig:topo}
\end{figure}

Having discussed the effective Hamiltonian of BPN, we consider its topological consequences. 
Topological characteristics regarding the band crossing at the type-II Dirac point is revealed by calculating the Zak phase. Here we consider the Zak phase of each band along the $k_y$ direction with fixed $k_x$,
$
    \gamma_n(k_x)=i\int_{-\pi}^{\pi} dk_y \; \langle u_n({\bf{k}}) \vert {\partial_{k_y}} \vert u_n({\bf{k}}) \rangle
$
where $u_n({\bf{k}})$ is the cell periodic function of the $n$-th Bloch state of Eq.~\ref{eq:tbH}. This quantity represents the Wannier center of the corresponding band along the $y$ direction and should be quantized by either 0 or $\pi$ (mod $2\pi$) due to the mirror symmetry of the system~\cite{PhysRevB.94.165164}. The Zak phases of the two bands that form the tilted Dirac cone are quantized as shown in Fig.~\ref{fig:topo} (a). Each band has a different quantized value that are exchanged at the crossing point. The discrete jump of the Zak phase makes the Berry phase along the circular path around the Dirac point be either $\pi$ or $-\pi$. The non-trivial Berry phase indicates pseudo-spin texture winding around the Dirac point as already conceived in Eq.~\ref{eq:typeII}. 
Moreover, once we translate half of the unitcell, the Zak phase of each band jumps from 0 to $\pi$ or vice versa compared to that from the original unitcell, because the Wannier center is translated by the same amount. Therefore, the domain boundaries between half-translated unitcells are the topological domain boundary where two bands with different Zak phases encounter and the topological states associated with the difference should occur. 
In 
Figs.~\ref{fig:topo} (b) and (c),
two different topological domain boundaries between half-translated unit cell structures can be constructed. In each domain boundary, topologically protected edge states that connect two opposite Dirac points emerge 
(Figs.~\ref{fig:topo}(d) and (e)).
The protected states on the domain boundaries are expected to show characteristic scanning tunneling microscopy images and spectroscopic signals as demonstrated recently for other 2D topological materials~\cite{Kim2020NL}.

In conclusion, we have shown that the new nonbenzenoid 2D carbon crystal  has interesting low energy states such as the zone-center vHS and near-critical tilted Dirac cones. These two states undergo topological transitions by merging themselves under strains.
We also expect that several anomalies related with the zone-center saddle point such as fluctuating magnetic states~\cite{Guinea1997PRL, Furukawa1998PRL, Honerkamp2001PRL, Yu2001PRB} 
as well as nontrivial magnetic responses~\cite{Giura1968PRL, Henrich1971PRL,Nikolaev2021PRB} 
could be investigated in the new carbon allotrope.

Y.-W.S. was supported by NRF of Korea (Grant No. 2017R1A5A1014862, SRC program: vdWMRC center) and KIAS individual Grant No. (CG031509).
H.J. was supported by NRF of Korea (Grant Nos. 2021M3H4A1A03054864, 2019R1A2C1010498, and 2017M3D1A1040833).
Computations were supported by the CAC of KIAS.


\begin{thebibliography}{43}%
\makeatletter
\providecommand \@ifxundefined [1]{%
 \@ifx{#1\undefined}
}%
\providecommand \@ifnum [1]{%
 \ifnum #1\expandafter \@firstoftwo
 \else \expandafter \@secondoftwo
 \fi
}%
\providecommand \@ifx [1]{%
 \ifx #1\expandafter \@firstoftwo
 \else \expandafter \@secondoftwo
 \fi
}%
\providecommand \natexlab [1]{#1}%
\providecommand \enquote  [1]{``#1''}%
\providecommand \bibnamefont  [1]{#1}%
\providecommand \bibfnamefont [1]{#1}%
\providecommand \citenamefont [1]{#1}%
\providecommand \href@noop [0]{\@secondoftwo}%
\providecommand \href [0]{\begingroup \@sanitize@url \@href}%
\providecommand \@href[1]{\@@startlink{#1}\@@href}%
\providecommand \@@href[1]{\endgroup#1\@@endlink}%
\providecommand \@sanitize@url [0]{\catcode `\\12\catcode `\$12\catcode
  `\&12\catcode `\#12\catcode `\^12\catcode `\_12\catcode `\%12\relax}%
\providecommand \@@startlink[1]{}%
\providecommand \@@endlink[0]{}%
\providecommand \url  [0]{\begingroup\@sanitize@url \@url }%
\providecommand \@url [1]{\endgroup\@href {#1}{\urlprefix }}%
\providecommand \urlprefix  [0]{URL }%
\providecommand \Eprint [0]{\href }%
\providecommand \doibase [0]{http://dx.doi.org/}%
\providecommand \selectlanguage [0]{\@gobble}%
\providecommand \bibinfo  [0]{\@secondoftwo}%
\providecommand \bibfield  [0]{\@secondoftwo}%
\providecommand \translation [1]{[#1]}%
\providecommand \BibitemOpen [0]{}%
\providecommand \bibitemStop [0]{}%
\providecommand \bibitemNoStop [0]{.\EOS\space}%
\providecommand \EOS [0]{\spacefactor3000\relax}%
\providecommand \BibitemShut  [1]{\csname bibitem#1\endcsname}%
\let\auto@bib@innerbib\@empty
\bibitem [{\citenamefont {Wallace}(1947)}]{Wallace1947PR}%
  \BibitemOpen
  \bibfield  {author} {\bibinfo {author} {\bibfnamefont {P.~R.}\ \bibnamefont
  {Wallace}},\ }\href {\doibase 10.1103/PhysRev.71.622} {\bibfield  {journal}
  {\bibinfo  {journal} {Phys. Rev.}\ }\textbf {\bibinfo {volume} {71}},\
  \bibinfo {pages} {622} (\bibinfo {year} {1947})}\BibitemShut {NoStop}%
\bibitem [{\citenamefont {Charlier}\ \emph {et~al.}(2007)\citenamefont
  {Charlier}, \citenamefont {Blase},\ and\ \citenamefont
  {Roche}}]{Charlier2007RMP}%
  \BibitemOpen
  \bibfield  {author} {\bibinfo {author} {\bibfnamefont {J.-C.}\ \bibnamefont
  {Charlier}}, \bibinfo {author} {\bibfnamefont {X.}~\bibnamefont {Blase}}, \
  and\ \bibinfo {author} {\bibfnamefont {S.}~\bibnamefont {Roche}},\ }\href
  {\doibase 10.1103/RevModPhys.79.677} {\bibfield  {journal} {\bibinfo
  {journal} {Rev. Mod. Phys.}\ }\textbf {\bibinfo {volume} {79}},\ \bibinfo
  {pages} {677} (\bibinfo {year} {2007})}\BibitemShut {NoStop}%
\bibitem [{\citenamefont {Castro~Neto}\ \emph {et~al.}(2009)\citenamefont
  {Castro~Neto}, \citenamefont {Guinea}, \citenamefont {Peres}, \citenamefont
  {Novoselov},\ and\ \citenamefont {Geim}}]{CastroNeto2009RMP}%
  \BibitemOpen
  \bibfield  {author} {\bibinfo {author} {\bibfnamefont {A.~H.}\ \bibnamefont
  {Castro~Neto}}, \bibinfo {author} {\bibfnamefont {F.}~\bibnamefont {Guinea}},
  \bibinfo {author} {\bibfnamefont {N.~M.~R.}\ \bibnamefont {Peres}}, \bibinfo
  {author} {\bibfnamefont {K.~S.}\ \bibnamefont {Novoselov}}, \ and\ \bibinfo
  {author} {\bibfnamefont {A.~K.}\ \bibnamefont {Geim}},\ }\href {\doibase
  10.1103/RevModPhys.81.109} {\bibfield  {journal} {\bibinfo  {journal} {Rev.
  Mod. Phys.}\ }\textbf {\bibinfo {volume} {81}},\ \bibinfo {pages} {109}
  (\bibinfo {year} {2009})}\BibitemShut {NoStop}%
\bibitem [{\citenamefont {Fan}\ \emph {et~al.}(2021)\citenamefont {Fan},
  \citenamefont {Yan}, \citenamefont {Tripp}, \citenamefont {Krej{\v c}{\'\i}},
  \citenamefont {Dimosthenous}, \citenamefont {Kachel}, \citenamefont {Chen},
  \citenamefont {Foster}, \citenamefont {Koert}, \citenamefont {Liljeroth},\
  and\ \citenamefont {Gottfried}}]{Fan2021Science}%
  \BibitemOpen
  \bibfield  {author} {\bibinfo {author} {\bibfnamefont {Q.}~\bibnamefont
  {Fan}}, \bibinfo {author} {\bibfnamefont {L.}~\bibnamefont {Yan}}, \bibinfo
  {author} {\bibfnamefont {M.~W.}\ \bibnamefont {Tripp}}, \bibinfo {author}
  {\bibfnamefont {O.}~\bibnamefont {Krej{\v c}{\'\i}}}, \bibinfo {author}
  {\bibfnamefont {S.}~\bibnamefont {Dimosthenous}}, \bibinfo {author}
  {\bibfnamefont {S.~R.}\ \bibnamefont {Kachel}}, \bibinfo {author}
  {\bibfnamefont {M.}~\bibnamefont {Chen}}, \bibinfo {author} {\bibfnamefont
  {A.~S.}\ \bibnamefont {Foster}}, \bibinfo {author} {\bibfnamefont
  {U.}~\bibnamefont {Koert}}, \bibinfo {author} {\bibfnamefont
  {P.}~\bibnamefont {Liljeroth}}, \ and\ \bibinfo {author} {\bibfnamefont
  {J.~M.}\ \bibnamefont {Gottfried}},\ }\href {\doibase
  10.1126/science.abg4509} {\bibfield  {journal} {\bibinfo  {journal}
  {Science}\ }\textbf {\bibinfo {volume} {372}},\ \bibinfo {pages} {852}
  (\bibinfo {year} {2021})}\BibitemShut {NoStop}%
\bibitem [{\citenamefont {Novoselov}\ \emph {et~al.}(2016)\citenamefont
  {Novoselov}, \citenamefont {Mishchenko}, \citenamefont {Carvalho},\ and\
  \citenamefont {{Castro Neto}}}]{Novoselov2016Science}%
  \BibitemOpen
  \bibfield  {author} {\bibinfo {author} {\bibfnamefont {K.~S.}\ \bibnamefont
  {Novoselov}}, \bibinfo {author} {\bibfnamefont {A.}~\bibnamefont
  {Mishchenko}}, \bibinfo {author} {\bibfnamefont {A.}~\bibnamefont
  {Carvalho}}, \ and\ \bibinfo {author} {\bibfnamefont {A.~H.}\ \bibnamefont
  {{Castro Neto}}},\ }\href {\doibase 10.1126/science.aac9439} {\bibfield
  {journal} {\bibinfo  {journal} {Science}\ }\textbf {\bibinfo {volume}
  {353}},\ \bibinfo {pages} {9439} (\bibinfo {year} {2016})}\BibitemShut
  {NoStop}%
\bibitem [{\citenamefont {Giannozzi}\ \emph {et~al.}(2009)\citenamefont
  {Giannozzi}, \citenamefont {Baroni}, \citenamefont {Bonini}, \citenamefont
  {Calandra}, \citenamefont {Car}, \citenamefont {Cavazzoni}, \citenamefont
  {Ceresoli}, \citenamefont {Chiarotti}, \citenamefont {Cococcioni},
  \citenamefont {Dabo}, \citenamefont {Corso}, \citenamefont {de~Gironcoli},
  \citenamefont {Fabris}, \citenamefont {Fratesi}, \citenamefont {Gebauer},
  \citenamefont {Gerstmann}, \citenamefont {Gougoussis}, \citenamefont
  {Kokalj}, \citenamefont {Lazzeri}, \citenamefont {Martin-Samos},
  \citenamefont {Marzari}, \citenamefont {Mauri}, \citenamefont {Mazzarello},
  \citenamefont {Paolini}, \citenamefont {Pasquarello}, \citenamefont
  {Paulatto}, \citenamefont {Sbraccia}, \citenamefont {Scandolo}, \citenamefont
  {Sclauzero}, \citenamefont {Seitsonen}, \citenamefont {Smogunov},
  \citenamefont {Umari},\ and\ \citenamefont {Wentzcovitch}}]{QE2009JPCM}%
  \BibitemOpen
  \bibfield  {author} {\bibinfo {author} {\bibfnamefont {P.}~\bibnamefont
  {Giannozzi}}, \bibinfo {author} {\bibfnamefont {S.}~\bibnamefont {Baroni}},
  \bibinfo {author} {\bibfnamefont {N.}~\bibnamefont {Bonini}}, \bibinfo
  {author} {\bibfnamefont {M.}~\bibnamefont {Calandra}}, \bibinfo {author}
  {\bibfnamefont {R.}~\bibnamefont {Car}}, \bibinfo {author} {\bibfnamefont
  {C.}~\bibnamefont {Cavazzoni}}, \bibinfo {author} {\bibfnamefont
  {D.}~\bibnamefont {Ceresoli}}, \bibinfo {author} {\bibfnamefont {G.~L.}\
  \bibnamefont {Chiarotti}}, \bibinfo {author} {\bibfnamefont {M.}~\bibnamefont
  {Cococcioni}}, \bibinfo {author} {\bibfnamefont {I.}~\bibnamefont {Dabo}},
  \bibinfo {author} {\bibfnamefont {A.~D.}\ \bibnamefont {Corso}}, \bibinfo
  {author} {\bibfnamefont {S.}~\bibnamefont {de~Gironcoli}}, \bibinfo {author}
  {\bibfnamefont {S.}~\bibnamefont {Fabris}}, \bibinfo {author} {\bibfnamefont
  {G.}~\bibnamefont {Fratesi}}, \bibinfo {author} {\bibfnamefont
  {R.}~\bibnamefont {Gebauer}}, \bibinfo {author} {\bibfnamefont
  {U.}~\bibnamefont {Gerstmann}}, \bibinfo {author} {\bibfnamefont
  {C.}~\bibnamefont {Gougoussis}}, \bibinfo {author} {\bibfnamefont
  {A.}~\bibnamefont {Kokalj}}, \bibinfo {author} {\bibfnamefont
  {M.}~\bibnamefont {Lazzeri}}, \bibinfo {author} {\bibfnamefont
  {L.}~\bibnamefont {Martin-Samos}}, \bibinfo {author} {\bibfnamefont
  {N.}~\bibnamefont {Marzari}}, \bibinfo {author} {\bibfnamefont
  {F.}~\bibnamefont {Mauri}}, \bibinfo {author} {\bibfnamefont
  {R.}~\bibnamefont {Mazzarello}}, \bibinfo {author} {\bibfnamefont
  {S.}~\bibnamefont {Paolini}}, \bibinfo {author} {\bibfnamefont
  {A.}~\bibnamefont {Pasquarello}}, \bibinfo {author} {\bibfnamefont
  {L.}~\bibnamefont {Paulatto}}, \bibinfo {author} {\bibfnamefont
  {C.}~\bibnamefont {Sbraccia}}, \bibinfo {author} {\bibfnamefont
  {S.}~\bibnamefont {Scandolo}}, \bibinfo {author} {\bibfnamefont
  {G.}~\bibnamefont {Sclauzero}}, \bibinfo {author} {\bibfnamefont {A.~P.}\
  \bibnamefont {Seitsonen}}, \bibinfo {author} {\bibfnamefont {A.}~\bibnamefont
  {Smogunov}}, \bibinfo {author} {\bibfnamefont {P.}~\bibnamefont {Umari}}, \
  and\ \bibinfo {author} {\bibfnamefont {R.~M.}\ \bibnamefont {Wentzcovitch}},\
  }\href {\doibase 10.1088/0953-8984/21/39/395502} {\bibfield  {journal}
  {\bibinfo  {journal} {J. Phys.: Condens. Matter}\ }\textbf {\bibinfo {volume}
  {21}},\ \bibinfo {pages} {395502} (\bibinfo {year} {2009})}\BibitemShut
  {NoStop}%
\bibitem [{\citenamefont {Perdew}\ \emph {et~al.}(1996)\citenamefont {Perdew},
  \citenamefont {Burke},\ and\ \citenamefont {Ernzerhof}}]{GGA1996PRL}%
  \BibitemOpen
  \bibfield  {author} {\bibinfo {author} {\bibfnamefont {J.~P.}\ \bibnamefont
  {Perdew}}, \bibinfo {author} {\bibfnamefont {K.}~\bibnamefont {Burke}}, \
  and\ \bibinfo {author} {\bibfnamefont {M.}~\bibnamefont {Ernzerhof}},\ }\href
  {\doibase 10.1103/PhysRevLett.77.3865} {\bibfield  {journal} {\bibinfo
  {journal} {Phys. Rev. Lett.}\ }\textbf {\bibinfo {volume} {77}},\ \bibinfo
  {pages} {3865} (\bibinfo {year} {1996})}\BibitemShut {NoStop}%
\bibitem [{\citenamefont {Garrity}\ \emph {et~al.}(2014)\citenamefont
  {Garrity}, \citenamefont {Bennett}, \citenamefont {Rabe},\ and\ \citenamefont
  {Vanderbilt}}]{GVRB2014CM}%
  \BibitemOpen
  \bibfield  {author} {\bibinfo {author} {\bibfnamefont {K.~F.}\ \bibnamefont
  {Garrity}}, \bibinfo {author} {\bibfnamefont {J.~W.}\ \bibnamefont
  {Bennett}}, \bibinfo {author} {\bibfnamefont {K.~M.}\ \bibnamefont {Rabe}}, \
  and\ \bibinfo {author} {\bibfnamefont {D.}~\bibnamefont {Vanderbilt}},\
  }\href {\doibase https://doi.org/10.1016/j.commatsci.2013.08.053} {\bibfield
  {journal} {\bibinfo  {journal} {Comput. Mater.}\ }\textbf {\bibinfo {volume}
  {81}},\ \bibinfo {pages} {446} (\bibinfo {year} {2014})}\BibitemShut
  {NoStop}%
\bibitem [{\citenamefont {Agapito}\ \emph {et~al.}(2015)\citenamefont
  {Agapito}, \citenamefont {Curtarolo},\ and\ \citenamefont
  {Buongiorno~Nardelli}}]{ACBN2015PRX}%
  \BibitemOpen
  \bibfield  {author} {\bibinfo {author} {\bibfnamefont {L.~A.}\ \bibnamefont
  {Agapito}}, \bibinfo {author} {\bibfnamefont {S.}~\bibnamefont {Curtarolo}},
  \ and\ \bibinfo {author} {\bibfnamefont {M.}~\bibnamefont
  {Buongiorno~Nardelli}},\ }\href {\doibase 10.1103/PhysRevX.5.011006}
  {\bibfield  {journal} {\bibinfo  {journal} {Phys. Rev. X}\ }\textbf {\bibinfo
  {volume} {5}},\ \bibinfo {pages} {011006} (\bibinfo {year}
  {2015})}\BibitemShut {NoStop}%
\bibitem [{\citenamefont {Lee}\ and\ \citenamefont {Son}(2020)}]{Lee2020PRR}%
  \BibitemOpen
  \bibfield  {author} {\bibinfo {author} {\bibfnamefont {S.-H.}\ \bibnamefont
  {Lee}}\ and\ \bibinfo {author} {\bibfnamefont {Y.-W.}\ \bibnamefont {Son}},\
  }\href {\doibase 10.1103/PhysRevResearch.2.043410} {\bibfield  {journal}
  {\bibinfo  {journal} {Phys. Rev. Research}\ }\textbf {\bibinfo {volume}
  {2}},\ \bibinfo {pages} {043410} (\bibinfo {year} {2020})}\BibitemShut
  {NoStop}%
\bibitem [{\citenamefont {Tancogne-Dejean}\ and\ \citenamefont
  {Rubio}(2020)}]{Rubio2020PRB}%
  \BibitemOpen
  \bibfield  {author} {\bibinfo {author} {\bibfnamefont {N.}~\bibnamefont
  {Tancogne-Dejean}}\ and\ \bibinfo {author} {\bibfnamefont {A.}~\bibnamefont
  {Rubio}},\ }\href {\doibase 10.1103/PhysRevB.102.155117} {\bibfield
  {journal} {\bibinfo  {journal} {Phys. Rev. B}\ }\textbf {\bibinfo {volume}
  {102}},\ \bibinfo {pages} {155117} (\bibinfo {year} {2020})}\BibitemShut
  {NoStop}%
\bibitem [{\citenamefont {{Campo Jr}}\ and\ \citenamefont
  {Cococcioni}(2010)}]{Cococcioni2010JPCM}%
  \BibitemOpen
  \bibfield  {author} {\bibinfo {author} {\bibfnamefont {V.~L.}\ \bibnamefont
  {{Campo Jr}}}\ and\ \bibinfo {author} {\bibfnamefont {M.}~\bibnamefont
  {Cococcioni}},\ }\href {\doibase 10.1088/0953-8984/22/5/055602} {\bibfield
  {journal} {\bibinfo  {journal} {J. Phys.: Condens. Matter}\ }\textbf
  {\bibinfo {volume} {22}},\ \bibinfo {pages} {055602} (\bibinfo {year}
  {2010})}\BibitemShut {NoStop}%
\bibitem [{\citenamefont {Timrov}\ \emph {et~al.}(2021)\citenamefont {Timrov},
  \citenamefont {Marzari},\ and\ \citenamefont {Cococcioni}}]{Timrov2021PRB}%
  \BibitemOpen
  \bibfield  {author} {\bibinfo {author} {\bibfnamefont {I.}~\bibnamefont
  {Timrov}}, \bibinfo {author} {\bibfnamefont {N.}~\bibnamefont {Marzari}}, \
  and\ \bibinfo {author} {\bibfnamefont {M.}~\bibnamefont {Cococcioni}},\
  }\href {\doibase 10.1103/PhysRevB.103.045141} {\bibfield  {journal} {\bibinfo
   {journal} {Phys. Rev. B}\ }\textbf {\bibinfo {volume} {103}},\ \bibinfo
  {pages} {045141} (\bibinfo {year} {2021})}\BibitemShut {NoStop}%
\bibitem [{\citenamefont {Hlubina}\ \emph {et~al.}(1997)\citenamefont
  {Hlubina}, \citenamefont {Sorella},\ and\ \citenamefont
  {Guinea}}]{Guinea1997PRL}%
  \BibitemOpen
  \bibfield  {author} {\bibinfo {author} {\bibfnamefont {R.}~\bibnamefont
  {Hlubina}}, \bibinfo {author} {\bibfnamefont {S.}~\bibnamefont {Sorella}}, \
  and\ \bibinfo {author} {\bibfnamefont {F.}~\bibnamefont {Guinea}},\ }\href
  {\doibase 10.1103/PhysRevLett.78.1343} {\bibfield  {journal} {\bibinfo
  {journal} {Phys. Rev. Lett.}\ }\textbf {\bibinfo {volume} {78}},\ \bibinfo
  {pages} {1343} (\bibinfo {year} {1997})}\BibitemShut {NoStop}%
\bibitem [{\citenamefont {Furukawa}\ \emph {et~al.}(1998)\citenamefont
  {Furukawa}, \citenamefont {Rice},\ and\ \citenamefont
  {Salmhofer}}]{Furukawa1998PRL}%
  \BibitemOpen
  \bibfield  {author} {\bibinfo {author} {\bibfnamefont {N.}~\bibnamefont
  {Furukawa}}, \bibinfo {author} {\bibfnamefont {T.~M.}\ \bibnamefont {Rice}},
  \ and\ \bibinfo {author} {\bibfnamefont {M.}~\bibnamefont {Salmhofer}},\
  }\href {\doibase 10.1103/PhysRevLett.81.3195} {\bibfield  {journal} {\bibinfo
   {journal} {Phys. Rev. Lett.}\ }\textbf {\bibinfo {volume} {81}},\ \bibinfo
  {pages} {3195} (\bibinfo {year} {1998})}\BibitemShut {NoStop}%
\bibitem [{\citenamefont {Honerkamp}\ and\ \citenamefont
  {Salmhofer}(2001)}]{Honerkamp2001PRL}%
  \BibitemOpen
  \bibfield  {author} {\bibinfo {author} {\bibfnamefont {C.}~\bibnamefont
  {Honerkamp}}\ and\ \bibinfo {author} {\bibfnamefont {M.}~\bibnamefont
  {Salmhofer}},\ }\href {\doibase 10.1103/PhysRevLett.87.187004} {\bibfield
  {journal} {\bibinfo  {journal} {Phys. Rev. Lett.}\ }\textbf {\bibinfo
  {volume} {87}},\ \bibinfo {pages} {187004} (\bibinfo {year}
  {2001})}\BibitemShut {NoStop}%
\bibitem [{\citenamefont {Irkhin}\ \emph {et~al.}(2001)\citenamefont {Irkhin},
  \citenamefont {Katanin},\ and\ \citenamefont {Katsnelson}}]{Yu2001PRB}%
  \BibitemOpen
  \bibfield  {author} {\bibinfo {author} {\bibfnamefont {V.~Y.}\ \bibnamefont
  {Irkhin}}, \bibinfo {author} {\bibfnamefont {A.~A.}\ \bibnamefont {Katanin}},
  \ and\ \bibinfo {author} {\bibfnamefont {M.~I.}\ \bibnamefont {Katsnelson}},\
  }\href {\doibase 10.1103/PhysRevB.64.165107} {\bibfield  {journal} {\bibinfo
  {journal} {Phys. Rev. B}\ }\textbf {\bibinfo {volume} {64}},\ \bibinfo
  {pages} {165107} (\bibinfo {year} {2001})}\BibitemShut {NoStop}%
\bibitem [{\citenamefont {Young}\ and\ \citenamefont
  {Kane}(2015)}]{Young2015PRL}%
  \BibitemOpen
  \bibfield  {author} {\bibinfo {author} {\bibfnamefont {S.~M.}\ \bibnamefont
  {Young}}\ and\ \bibinfo {author} {\bibfnamefont {C.~L.}\ \bibnamefont
  {Kane}},\ }\href {\doibase 10.1103/PhysRevLett.115.126803} {\bibfield
  {journal} {\bibinfo  {journal} {Phys. Rev. Lett.}\ }\textbf {\bibinfo
  {volume} {115}},\ \bibinfo {pages} {126803} (\bibinfo {year}
  {2015})}\BibitemShut {NoStop}%
\bibitem [{\citenamefont {Kim}\ \emph {et~al.}(2015)\citenamefont {Kim},
  \citenamefont {Wieder}, \citenamefont {Kane},\ and\ \citenamefont
  {Rappe}}]{Kim2015PRL}%
  \BibitemOpen
  \bibfield  {author} {\bibinfo {author} {\bibfnamefont {Y.}~\bibnamefont
  {Kim}}, \bibinfo {author} {\bibfnamefont {B.~J.}\ \bibnamefont {Wieder}},
  \bibinfo {author} {\bibfnamefont {C.~L.}\ \bibnamefont {Kane}}, \ and\
  \bibinfo {author} {\bibfnamefont {A.~M.}\ \bibnamefont {Rappe}},\ }\href
  {\doibase 10.1103/PhysRevLett.115.036806} {\bibfield  {journal} {\bibinfo
  {journal} {Phys. Rev. Lett.}\ }\textbf {\bibinfo {volume} {115}},\ \bibinfo
  {pages} {036806} (\bibinfo {year} {2015})}\BibitemShut {NoStop}%
\bibitem [{\citenamefont {Soluyanov}\ \emph {et~al.}(2015)\citenamefont
  {Soluyanov}, \citenamefont {Gresch}, \citenamefont {Wang}, \citenamefont
  {Wu}, \citenamefont {Troyer}, \citenamefont {Dai},\ and\ \citenamefont
  {Bernevig}}]{Bernevig2015Nature}%
  \BibitemOpen
  \bibfield  {author} {\bibinfo {author} {\bibfnamefont {A.~A.}\ \bibnamefont
  {Soluyanov}}, \bibinfo {author} {\bibfnamefont {D.}~\bibnamefont {Gresch}},
  \bibinfo {author} {\bibfnamefont {Z.}~\bibnamefont {Wang}}, \bibinfo {author}
  {\bibfnamefont {Q.}~\bibnamefont {Wu}}, \bibinfo {author} {\bibfnamefont
  {M.}~\bibnamefont {Troyer}}, \bibinfo {author} {\bibfnamefont
  {X.}~\bibnamefont {Dai}}, \ and\ \bibinfo {author} {\bibfnamefont {B.~A.}\
  \bibnamefont {Bernevig}},\ }\href {\doibase 10.1038/nature15768} {\bibfield
  {journal} {\bibinfo  {journal} {Nature}\ }\textbf {\bibinfo {volume} {527}},\
  \bibinfo {pages} {495} (\bibinfo {year} {2015})}\BibitemShut {NoStop}%
\bibitem [{\citenamefont {Yan}\ \emph {et~al.}(2017)\citenamefont {Yan},
  \citenamefont {Huang}, \citenamefont {Zhang}, \citenamefont {Wang},
  \citenamefont {Yao}, \citenamefont {Deng}, \citenamefont {Wan}, \citenamefont
  {Zhang}, \citenamefont {Arita}, \citenamefont {Yang}, \citenamefont {Sun},
  \citenamefont {Yao}, \citenamefont {Wu}, \citenamefont {Fan}, \citenamefont
  {Duan},\ and\ \citenamefont {Zhou}}]{Shuyun2017NatComm}%
  \BibitemOpen
  \bibfield  {author} {\bibinfo {author} {\bibfnamefont {M.}~\bibnamefont
  {Yan}}, \bibinfo {author} {\bibfnamefont {H.}~\bibnamefont {Huang}}, \bibinfo
  {author} {\bibfnamefont {K.}~\bibnamefont {Zhang}}, \bibinfo {author}
  {\bibfnamefont {E.}~\bibnamefont {Wang}}, \bibinfo {author} {\bibfnamefont
  {W.}~\bibnamefont {Yao}}, \bibinfo {author} {\bibfnamefont {K.}~\bibnamefont
  {Deng}}, \bibinfo {author} {\bibfnamefont {G.}~\bibnamefont {Wan}}, \bibinfo
  {author} {\bibfnamefont {H.}~\bibnamefont {Zhang}}, \bibinfo {author}
  {\bibfnamefont {M.}~\bibnamefont {Arita}}, \bibinfo {author} {\bibfnamefont
  {H.}~\bibnamefont {Yang}}, \bibinfo {author} {\bibfnamefont {Z.}~\bibnamefont
  {Sun}}, \bibinfo {author} {\bibfnamefont {H.}~\bibnamefont {Yao}}, \bibinfo
  {author} {\bibfnamefont {Y.}~\bibnamefont {Wu}}, \bibinfo {author}
  {\bibfnamefont {S.}~\bibnamefont {Fan}}, \bibinfo {author} {\bibfnamefont
  {W.}~\bibnamefont {Duan}}, \ and\ \bibinfo {author} {\bibfnamefont
  {S.}~\bibnamefont {Zhou}},\ }\href {\doibase 10.1038/s41467-017-00280-6}
  {\bibfield  {journal} {\bibinfo  {journal} {Nat. Comm.}\ }\textbf {\bibinfo
  {volume} {8}},\ \bibinfo {pages} {257} (\bibinfo {year} {2017})}\BibitemShut
  {NoStop}%
\bibitem [{\citenamefont {Armitage}\ \emph {et~al.}(2018)\citenamefont
  {Armitage}, \citenamefont {Mele},\ and\ \citenamefont
  {Vishwanath}}]{Ashvin2018RMP}%
  \BibitemOpen
  \bibfield  {author} {\bibinfo {author} {\bibfnamefont {N.~P.}\ \bibnamefont
  {Armitage}}, \bibinfo {author} {\bibfnamefont {E.~J.}\ \bibnamefont {Mele}},
  \ and\ \bibinfo {author} {\bibfnamefont {A.}~\bibnamefont {Vishwanath}},\
  }\href {\doibase 10.1103/RevModPhys.90.015001} {\bibfield  {journal}
  {\bibinfo  {journal} {Rev. Mod. Phys.}\ }\textbf {\bibinfo {volume} {90}},\
  \bibinfo {pages} {015001} (\bibinfo {year} {2018})}\BibitemShut {NoStop}%
\bibitem [{\citenamefont {Baroni}\ \emph {et~al.}(2001)\citenamefont {Baroni},
  \citenamefont {de~Gironcoli}, \citenamefont {Dal~Corso},\ and\ \citenamefont
  {Giannozzi}}]{DFPT2001RMP}%
  \BibitemOpen
  \bibfield  {author} {\bibinfo {author} {\bibfnamefont {S.}~\bibnamefont
  {Baroni}}, \bibinfo {author} {\bibfnamefont {S.}~\bibnamefont
  {de~Gironcoli}}, \bibinfo {author} {\bibfnamefont {A.}~\bibnamefont
  {Dal~Corso}}, \ and\ \bibinfo {author} {\bibfnamefont {P.}~\bibnamefont
  {Giannozzi}},\ }\href {\doibase 10.1103/RevModPhys.73.515} {\bibfield
  {journal} {\bibinfo  {journal} {Rev. Mod. Phys.}\ }\textbf {\bibinfo {volume}
  {73}},\ \bibinfo {pages} {515} (\bibinfo {year} {2001})}\BibitemShut
  {NoStop}%
\bibitem [{\citenamefont {Veeravenkata}\ and\ \citenamefont
  {Jain}(2021)}]{Jain2021Carbon}%
  \BibitemOpen
  \bibfield  {author} {\bibinfo {author} {\bibfnamefont {H.~P.}\ \bibnamefont
  {Veeravenkata}}\ and\ \bibinfo {author} {\bibfnamefont {A.}~\bibnamefont
  {Jain}},\ }\href {\doibase https://doi.org/10.1016/j.carbon.2021.07.078}
  {\bibfield  {journal} {\bibinfo  {journal} {Carbon}\ }\textbf {\bibinfo
  {volume} {183}},\ \bibinfo {pages} {893} (\bibinfo {year}
  {2021})}\BibitemShut {NoStop}%
\bibitem [{\citenamefont {Volovik}\ and\ \citenamefont
  {Zhang}(2017)}]{Volovik2017JLTP}%
  \BibitemOpen
  \bibfield  {author} {\bibinfo {author} {\bibfnamefont {G.~E.}\ \bibnamefont
  {Volovik}}\ and\ \bibinfo {author} {\bibfnamefont {K.}~\bibnamefont
  {Zhang}},\ }\href {https://doi.org/10.1007/s10909-017-1817-8} {\bibfield
  {journal} {\bibinfo  {journal} {J. of Low Temp. Phys.}\ }\textbf {\bibinfo
  {volume} {189}},\ \bibinfo {pages} {276} (\bibinfo {year}
  {2017})}\BibitemShut {NoStop}%
\bibitem [{\citenamefont {McCormick}\ \emph {et~al.}(2017)\citenamefont
  {McCormick}, \citenamefont {Kimchi},\ and\ \citenamefont
  {Trivedi}}]{McCormick2017PRB}%
  \BibitemOpen
  \bibfield  {author} {\bibinfo {author} {\bibfnamefont {T.~M.}\ \bibnamefont
  {McCormick}}, \bibinfo {author} {\bibfnamefont {I.}~\bibnamefont {Kimchi}}, \
  and\ \bibinfo {author} {\bibfnamefont {N.}~\bibnamefont {Trivedi}},\ }\href
  {\doibase 10.1103/PhysRevB.95.075133} {\bibfield  {journal} {\bibinfo
  {journal} {Phys. Rev. B}\ }\textbf {\bibinfo {volume} {95}},\ \bibinfo
  {pages} {075133} (\bibinfo {year} {2017})}\BibitemShut {NoStop}%
\bibitem [{\citenamefont {Kim}\ and\ \citenamefont {Son}(2021)}]{Kim2021PRB}%
  \BibitemOpen
  \bibfield  {author} {\bibinfo {author} {\bibfnamefont {S.}~\bibnamefont
  {Kim}}\ and\ \bibinfo {author} {\bibfnamefont {Y.-W.}\ \bibnamefont {Son}},\
  }\href {\doibase 10.1103/PhysRevB.104.045426} {\bibfield  {journal} {\bibinfo
   {journal} {Phys. Rev. B}\ }\textbf {\bibinfo {volume} {104}},\ \bibinfo
  {pages} {045426} (\bibinfo {year} {2021})}\BibitemShut {NoStop}%
\bibitem [{\citenamefont {Wehling}\ \emph {et~al.}(2011)\citenamefont
  {Wehling}, \citenamefont {\ifmmode \mbox{\c{S}}\else \c{S}\fi{}a\ifmmode
  \mbox{\c{s}}\else \c{s}\fi{}\ifmmode \imath \else \i
  \fi{}o\ifmmode~\breve{g}\else \u{g}\fi{}lu}, \citenamefont {Friedrich},
  \citenamefont {Lichtenstein}, \citenamefont {Katsnelson},\ and\ \citenamefont
  {Bl\"ugel}}]{Wehling2011PRL}%
  \BibitemOpen
  \bibfield  {author} {\bibinfo {author} {\bibfnamefont {T.~O.}\ \bibnamefont
  {Wehling}}, \bibinfo {author} {\bibfnamefont {E.}~\bibnamefont {\ifmmode
  \mbox{\c{S}}\else \c{S}\fi{}a\ifmmode \mbox{\c{s}}\else \c{s}\fi{}\ifmmode
  \imath \else \i \fi{}o\ifmmode~\breve{g}\else \u{g}\fi{}lu}}, \bibinfo
  {author} {\bibfnamefont {C.}~\bibnamefont {Friedrich}}, \bibinfo {author}
  {\bibfnamefont {A.~I.}\ \bibnamefont {Lichtenstein}}, \bibinfo {author}
  {\bibfnamefont {M.~I.}\ \bibnamefont {Katsnelson}}, \ and\ \bibinfo {author}
  {\bibfnamefont {S.}~\bibnamefont {Bl\"ugel}},\ }\href {\doibase
  10.1103/PhysRevLett.106.236805} {\bibfield  {journal} {\bibinfo  {journal}
  {Phys. Rev. Lett.}\ }\textbf {\bibinfo {volume} {106}},\ \bibinfo {pages}
  {236805} (\bibinfo {year} {2011})}\BibitemShut {NoStop}%
\bibitem [{\citenamefont {Ziletti}\ \emph {et~al.}(2015)\citenamefont
  {Ziletti}, \citenamefont {Huang}, \citenamefont {Coker},\ and\ \citenamefont
  {Lin}}]{Lin2015PRB}%
  \BibitemOpen
  \bibfield  {author} {\bibinfo {author} {\bibfnamefont {A.}~\bibnamefont
  {Ziletti}}, \bibinfo {author} {\bibfnamefont {S.~M.}\ \bibnamefont {Huang}},
  \bibinfo {author} {\bibfnamefont {D.~F.}\ \bibnamefont {Coker}}, \ and\
  \bibinfo {author} {\bibfnamefont {H.}~\bibnamefont {Lin}},\ }\href {\doibase
  10.1103/PhysRevB.92.085423} {\bibfield  {journal} {\bibinfo  {journal} {Phys.
  Rev. B}\ }\textbf {\bibinfo {volume} {92}},\ \bibinfo {pages} {085423}
  (\bibinfo {year} {2015})}\BibitemShut {NoStop}%
\bibitem [{\citenamefont {Gong}\ \emph {et~al.}(2017)\citenamefont {Gong},
  \citenamefont {Li}, \citenamefont {Li}, \citenamefont {Ji}, \citenamefont
  {Stern}, \citenamefont {Xia}, \citenamefont {Cao}, \citenamefont {Bao},
  \citenamefont {Wang}, \citenamefont {Wang}, \citenamefont {Qiu},
  \citenamefont {Cava}, \citenamefont {Louie}, \citenamefont {Xia},\ and\
  \citenamefont {Zhang}}]{Gong2017Nature}%
  \BibitemOpen
  \bibfield  {author} {\bibinfo {author} {\bibfnamefont {C.}~\bibnamefont
  {Gong}}, \bibinfo {author} {\bibfnamefont {L.}~\bibnamefont {Li}}, \bibinfo
  {author} {\bibfnamefont {Z.}~\bibnamefont {Li}}, \bibinfo {author}
  {\bibfnamefont {H.}~\bibnamefont {Ji}}, \bibinfo {author} {\bibfnamefont
  {A.}~\bibnamefont {Stern}}, \bibinfo {author} {\bibfnamefont
  {Y.}~\bibnamefont {Xia}}, \bibinfo {author} {\bibfnamefont {T.}~\bibnamefont
  {Cao}}, \bibinfo {author} {\bibfnamefont {W.}~\bibnamefont {Bao}}, \bibinfo
  {author} {\bibfnamefont {C.}~\bibnamefont {Wang}}, \bibinfo {author}
  {\bibfnamefont {Y.}~\bibnamefont {Wang}}, \bibinfo {author} {\bibfnamefont
  {Z.~Q.}\ \bibnamefont {Qiu}}, \bibinfo {author} {\bibfnamefont {R.~J.}\
  \bibnamefont {Cava}}, \bibinfo {author} {\bibfnamefont {S.~G.}\ \bibnamefont
  {Louie}}, \bibinfo {author} {\bibfnamefont {J.}~\bibnamefont {Xia}}, \ and\
  \bibinfo {author} {\bibfnamefont {X.}~\bibnamefont {Zhang}},\ }\href
  {\doibase 10.1038/nature22060} {\bibfield  {journal} {\bibinfo  {journal}
  {Nature}\ }\textbf {\bibinfo {volume} {546}},\ \bibinfo {pages} {265}
  (\bibinfo {year} {2017})}\BibitemShut {NoStop}%
\bibitem [{\citenamefont {Kim}\ \emph {et~al.}(2019)\citenamefont {Kim},
  \citenamefont {Lim}, \citenamefont {Lee}, \citenamefont {Lee}, \citenamefont
  {Kim}, \citenamefont {Park}, \citenamefont {Jeon}, \citenamefont {Park},
  \citenamefont {Park},\ and\ \citenamefont {Cheong}}]{Kim2019NatComm}%
  \BibitemOpen
  \bibfield  {author} {\bibinfo {author} {\bibfnamefont {K.}~\bibnamefont
  {Kim}}, \bibinfo {author} {\bibfnamefont {S.~Y.}\ \bibnamefont {Lim}},
  \bibinfo {author} {\bibfnamefont {J.-U.}\ \bibnamefont {Lee}}, \bibinfo
  {author} {\bibfnamefont {S.}~\bibnamefont {Lee}}, \bibinfo {author}
  {\bibfnamefont {T.~Y.}\ \bibnamefont {Kim}}, \bibinfo {author} {\bibfnamefont
  {K.}~\bibnamefont {Park}}, \bibinfo {author} {\bibfnamefont {G.~S.}\
  \bibnamefont {Jeon}}, \bibinfo {author} {\bibfnamefont {C.-H.}\ \bibnamefont
  {Park}}, \bibinfo {author} {\bibfnamefont {J.-G.}\ \bibnamefont {Park}}, \
  and\ \bibinfo {author} {\bibfnamefont {H.}~\bibnamefont {Cheong}},\ }\href
  {\doibase 10.1038/s41467-018-08284-6} {\bibfield  {journal} {\bibinfo
  {journal} {Nature Comm.}\ }\textbf {\bibinfo {volume} {10}},\ \bibinfo
  {pages} {345} (\bibinfo {year} {2019})}\BibitemShut {NoStop}%
\bibitem [{\citenamefont {Pereira}\ and\ \citenamefont
  {Castro~Neto}(2009)}]{Pereira2009PRL}%
  \BibitemOpen
  \bibfield  {author} {\bibinfo {author} {\bibfnamefont {V.~M.}\ \bibnamefont
  {Pereira}}\ and\ \bibinfo {author} {\bibfnamefont {A.~H.}\ \bibnamefont
  {Castro~Neto}},\ }\href {\doibase 10.1103/PhysRevLett.103.046801} {\bibfield
  {journal} {\bibinfo  {journal} {Phys. Rev. Lett.}\ }\textbf {\bibinfo
  {volume} {103}},\ \bibinfo {pages} {046801} (\bibinfo {year}
  {2009})}\BibitemShut {NoStop}%
\bibitem [{\citenamefont {Choi}\ \emph {et~al.}(2010)\citenamefont {Choi},
  \citenamefont {Jhi},\ and\ \citenamefont {Son}}]{Choi2010PRB}%
  \BibitemOpen
  \bibfield  {author} {\bibinfo {author} {\bibfnamefont {S.-M.}\ \bibnamefont
  {Choi}}, \bibinfo {author} {\bibfnamefont {S.-H.}\ \bibnamefont {Jhi}}, \
  and\ \bibinfo {author} {\bibfnamefont {Y.-W.}\ \bibnamefont {Son}},\ }\href
  {\doibase 10.1103/PhysRevB.81.081407} {\bibfield  {journal} {\bibinfo
  {journal} {Phys. Rev. B}\ }\textbf {\bibinfo {volume} {81}},\ \bibinfo
  {pages} {081407} (\bibinfo {year} {2010})}\BibitemShut {NoStop}%
\bibitem [{\citenamefont {Baik}\ \emph {et~al.}(2015)\citenamefont {Baik},
  \citenamefont {Kim}, \citenamefont {Yi},\ and\ \citenamefont
  {Choi}}]{baik2015NL}%
  \BibitemOpen
  \bibfield  {author} {\bibinfo {author} {\bibfnamefont {S.~S.}\ \bibnamefont
  {Baik}}, \bibinfo {author} {\bibfnamefont {K.~S.}\ \bibnamefont {Kim}},
  \bibinfo {author} {\bibfnamefont {Y.}~\bibnamefont {Yi}}, \ and\ \bibinfo
  {author} {\bibfnamefont {H.~J.}\ \bibnamefont {Choi}},\ }\href {\doibase
  10.1021/acs.nanolett.5b04106} {\bibfield  {journal} {\bibinfo  {journal}
  {Nano Lett.}\ }\textbf {\bibinfo {volume} {15}},\ \bibinfo {pages} {7788}
  (\bibinfo {year} {2015})}\BibitemShut {NoStop}%
\bibitem [{\citenamefont {Doh}\ and\ \citenamefont {Choi}(2017)}]{Doh2017Mat}%
  \BibitemOpen
  \bibfield  {author} {\bibinfo {author} {\bibfnamefont {H.}~\bibnamefont
  {Doh}}\ and\ \bibinfo {author} {\bibfnamefont {H.~J.}\ \bibnamefont {Choi}},\
  }\href {\doibase 10.1088/2053-1583/aa6835} {\bibfield  {journal} {\bibinfo
  {journal} {2D Mater.}\ }\textbf {\bibinfo {volume} {4}},\ \bibinfo {pages}
  {025071} (\bibinfo {year} {2017})}\BibitemShut {NoStop}%
\bibitem [{\citenamefont {Murakami}\ \emph {et~al.}(2007)\citenamefont
  {Murakami}, \citenamefont {Iso}, \citenamefont {Avishai}, \citenamefont
  {Onoda},\ and\ \citenamefont {Nagaosa}}]{Murakami2007PRB}%
  \BibitemOpen
  \bibfield  {author} {\bibinfo {author} {\bibfnamefont {S.}~\bibnamefont
  {Murakami}}, \bibinfo {author} {\bibfnamefont {S.}~\bibnamefont {Iso}},
  \bibinfo {author} {\bibfnamefont {Y.}~\bibnamefont {Avishai}}, \bibinfo
  {author} {\bibfnamefont {M.}~\bibnamefont {Onoda}}, \ and\ \bibinfo {author}
  {\bibfnamefont {N.}~\bibnamefont {Nagaosa}},\ }\href {\doibase
  10.1103/PhysRevB.76.205304} {\bibfield  {journal} {\bibinfo  {journal} {Phys.
  Rev. B}\ }\textbf {\bibinfo {volume} {76}},\ \bibinfo {pages} {205304}
  (\bibinfo {year} {2007})}\BibitemShut {NoStop}%
\bibitem [{\citenamefont {Son}\ \emph {et~al.}(2011)\citenamefont {Son},
  \citenamefont {Choi}, \citenamefont {Hong}, \citenamefont {Woo},\ and\
  \citenamefont {Jhi}}]{Son2011PRB}%
  \BibitemOpen
  \bibfield  {author} {\bibinfo {author} {\bibfnamefont {Y.-W.}\ \bibnamefont
  {Son}}, \bibinfo {author} {\bibfnamefont {S.-M.}\ \bibnamefont {Choi}},
  \bibinfo {author} {\bibfnamefont {Y.~P.}\ \bibnamefont {Hong}}, \bibinfo
  {author} {\bibfnamefont {S.}~\bibnamefont {Woo}}, \ and\ \bibinfo {author}
  {\bibfnamefont {S.-H.}\ \bibnamefont {Jhi}},\ }\href {\doibase
  10.1103/PhysRevB.84.155410} {\bibfield  {journal} {\bibinfo  {journal} {Phys.
  Rev. B}\ }\textbf {\bibinfo {volume} {84}},\ \bibinfo {pages} {155410}
  (\bibinfo {year} {2011})}\BibitemShut {NoStop}%
\bibitem [{\citenamefont {Dietl}\ \emph {et~al.}(2008)\citenamefont {Dietl},
  \citenamefont {Pi\'echon},\ and\ \citenamefont {Montambaux}}]{Dietl2008PRL}%
  \BibitemOpen
  \bibfield  {author} {\bibinfo {author} {\bibfnamefont {P.}~\bibnamefont
  {Dietl}}, \bibinfo {author} {\bibfnamefont {F.}~\bibnamefont {Pi\'echon}}, \
  and\ \bibinfo {author} {\bibfnamefont {G.}~\bibnamefont {Montambaux}},\
  }\href {\doibase 10.1103/PhysRevLett.100.236405} {\bibfield  {journal}
  {\bibinfo  {journal} {Phys. Rev. Lett.}\ }\textbf {\bibinfo {volume} {100}},\
  \bibinfo {pages} {236405} (\bibinfo {year} {2008})}\BibitemShut {NoStop}%
\bibitem [{\citenamefont {Lau}\ \emph {et~al.}(2016)\citenamefont {Lau},
  \citenamefont {van~den Brink},\ and\ \citenamefont
  {Ortix}}]{PhysRevB.94.165164}%
  \BibitemOpen
  \bibfield  {author} {\bibinfo {author} {\bibfnamefont {A.}~\bibnamefont
  {Lau}}, \bibinfo {author} {\bibfnamefont {J.}~\bibnamefont {van~den Brink}},
  \ and\ \bibinfo {author} {\bibfnamefont {C.}~\bibnamefont {Ortix}},\ }\href
  {\doibase 10.1103/PhysRevB.94.165164} {\bibfield  {journal} {\bibinfo
  {journal} {Phys. Rev. B}\ }\textbf {\bibinfo {volume} {94}},\ \bibinfo
  {pages} {165164} (\bibinfo {year} {2016})}\BibitemShut {NoStop}%
\bibitem [{\citenamefont {Kim}\ \emph {et~al.}(2020)\citenamefont {Kim},
  \citenamefont {Kang}, \citenamefont {Kim}, \citenamefont {Chae},
  \citenamefont {Cho}, \citenamefont {Ko}, \citenamefont {Jeon}, \citenamefont
  {Kang}, \citenamefont {Yang}, \citenamefont {Kim}, \citenamefont {Park},
  \citenamefont {Hwang}, \citenamefont {Kwon},\ and\ \citenamefont
  {Son}}]{Kim2020NL}%
  \BibitemOpen
  \bibfield  {author} {\bibinfo {author} {\bibfnamefont {H.~W.}\ \bibnamefont
  {Kim}}, \bibinfo {author} {\bibfnamefont {S.-H.}\ \bibnamefont {Kang}},
  \bibinfo {author} {\bibfnamefont {H.-J.}\ \bibnamefont {Kim}}, \bibinfo
  {author} {\bibfnamefont {K.}~\bibnamefont {Chae}}, \bibinfo {author}
  {\bibfnamefont {S.}~\bibnamefont {Cho}}, \bibinfo {author} {\bibfnamefont
  {W.}~\bibnamefont {Ko}}, \bibinfo {author} {\bibfnamefont {S.}~\bibnamefont
  {Jeon}}, \bibinfo {author} {\bibfnamefont {S.~H.}\ \bibnamefont {Kang}},
  \bibinfo {author} {\bibfnamefont {H.}~\bibnamefont {Yang}}, \bibinfo {author}
  {\bibfnamefont {S.~W.}\ \bibnamefont {Kim}}, \bibinfo {author} {\bibfnamefont
  {S.}~\bibnamefont {Park}}, \bibinfo {author} {\bibfnamefont {S.}~\bibnamefont
  {Hwang}}, \bibinfo {author} {\bibfnamefont {Y.-K.}\ \bibnamefont {Kwon}}, \
  and\ \bibinfo {author} {\bibfnamefont {Y.-W.}\ \bibnamefont {Son}},\ }\href
  {\doibase 10.1021/acs.nanolett.0c01756} {\bibfield  {journal} {\bibinfo
  {journal} {Nano Letters}\ }\textbf {\bibinfo {volume} {20}},\ \bibinfo
  {pages} {5837} (\bibinfo {year} {2020})}\BibitemShut {NoStop}%
\bibitem [{\citenamefont {Giura}\ and\ \citenamefont
  {Wanderlingh}(1968)}]{Giura1968PRL}%
  \BibitemOpen
  \bibfield  {author} {\bibinfo {author} {\bibfnamefont {M.}~\bibnamefont
  {Giura}}\ and\ \bibinfo {author} {\bibfnamefont {F.}~\bibnamefont
  {Wanderlingh}},\ }\href {\doibase 10.1103/PhysRevLett.20.445} {\bibfield
  {journal} {\bibinfo  {journal} {Phys. Rev. Lett.}\ }\textbf {\bibinfo
  {volume} {20}},\ \bibinfo {pages} {445} (\bibinfo {year} {1968})}\BibitemShut
  {NoStop}%
\bibitem [{\citenamefont {Henrich}(1971)}]{Henrich1971PRL}%
  \BibitemOpen
  \bibfield  {author} {\bibinfo {author} {\bibfnamefont {V.~E.}\ \bibnamefont
  {Henrich}},\ }\href {\doibase 10.1103/PhysRevLett.26.891} {\bibfield
  {journal} {\bibinfo  {journal} {Phys. Rev. Lett.}\ }\textbf {\bibinfo
  {volume} {26}},\ \bibinfo {pages} {891} (\bibinfo {year} {1971})}\BibitemShut
  {NoStop}%
\bibitem [{\citenamefont {Nikolaev}(2021)}]{Nikolaev2021PRB}%
  \BibitemOpen
  \bibfield  {author} {\bibinfo {author} {\bibfnamefont {A.~V.}\ \bibnamefont
  {Nikolaev}},\ }\href {\doibase 10.1103/PhysRevB.104.035419} {\bibfield
  {journal} {\bibinfo  {journal} {Phys. Rev. B}\ }\textbf {\bibinfo {volume}
  {104}},\ \bibinfo {pages} {035419} (\bibinfo {year} {2021})}\BibitemShut
  {NoStop}%
\end{thebibliography}
\end{document}